\documentclass{article}
\usepackage{spconf,amsmath,epsfig,graphicx}
\usepackage{subcaption}
\usepackage{caption}
\usepackage[breaklinks=true]{hyperref}
\usepackage{breakcites}
\usepackage{enumitem}
\usepackage{pbox}

\begin{document}\sloppy

\def\x{{\mathbf x}}
\def\L{{\cal L}}
\def\eg{\textit{e.g.}}
\def\ie{\textit{i.e.}}
\def\Eg{\textit{E.g.}}
\def\etal{\textit{et al.}}
\def\etc{\textit{etc.}}

\title{An Eye-tracking Dataset for Animal-centric PASCAL Object Classes}
%
\name{Syed Omer Gilani$^1$, Ramanathan Subramanian$^2$, Yan Yan$^3$, David Melcher$^4$, Nicu Sebe$^3$, Stefan Winkler}
\address{$^1$ SMME, National University of Sciences \& Technology, Islamabad, Pakistan \\
      $^2$ Advanced Digital Sciences Center, University of Illinois at Urbana-Champaign, Singapore \\
			$^3$ Department of Computer Science and Information Engineering, University of Trento, Italy \\
			$^4$ Department of Cognitive Sciences, University of Trento, Italy \\
			$^1$omer@smme.nust.edu.pk,$^2${subramanian.r,stefan.winkler}@adsc.com.sg}

\maketitle

\begin{abstract}
We present \textbf{PET}-- the Pascal animal classes Eye Tracking database. Our database comprises eye movement recordings compiled from forty users for the \textit{bird}, \textit{cat}, \textit{cow}, \textit{dog}, \textit{horse} and \textit{sheep} {trainval} sets from the VOC 2012 image set. Different from recent eye-tracking databases such as~\cite{kiwon_cvpr13_gaze,PapadopoulosCKF14}, a salient aspect of PET is that it contains eye movements recorded for both the \textit{\textbf{free-viewing}} and \textit{\textbf{visual search}} task conditions. While some differences in terms of overall gaze behavior and scanning patterns are observed between the two conditions, a very similar number of fixations are observed on target objects for both conditions. As a utility application, we show how feature pooling around fixated locations enables enhanced (animal) object classification accuracy.  
\end{abstract}

\section{Introduction}

The notion of utilizing user inputs, in the form  of explicit textual tags or implicit signals such as eye movements~\cite{RamanathanKSKC10,subramanian2014emotion} or brain responses~\cite{khomami2013user}, to interactively guide automated scene understanding algorithms has gained in popularity recently. With reference to the use of eye movements, a number of algorithms that learn from eye \textit{fixations} and \textit{saccades} to infer salient image regions and objects have been proposed in literature~\cite{Itti_Koch01nrn,subramanian2011can,borji2012boosting}. Fixations denote stationary phases during scene viewing during which humans encode visual information, while saccades denote ballistic eye movements to encode information pertaining to different scene regions. 

Of late, some works have specifically looked at utilizing eye fixations for facilitating object detection on the Pascal VOC image set~\cite{pascalvoc2012}. Two notable works in this respect are~\cite{kiwon_cvpr13_gaze} and~\cite{ PapadopoulosCKF14}. Both these works are based on the assumption that eye movements are concentrated on the salient object(s), and therefore, can enable (i) implicit and fast annotation of objects for model training, and (ii) enhanced  object detection performance. 

However, there is an important difference between the manner in which~\cite{kiwon_cvpr13_gaze} and~\cite{PapadopoulosCKF14} compile user eye movements. In~\cite{kiwon_cvpr13_gaze}, the authors record eye movements under a free-viewing paradigm hypothesizing that natural gaze patterns are capable of directly revealing salient image content. In contrast,~\cite{PapadopoulosCKF14} employs a visual search paradigm to compile eye movements based on the argument that free-viewing may not provide optimal data for training object detectors. While the impact of the task-on-hand on eye movements has been known and studied for long~\cite{Yarbus1967,deangelus2009top,tatler2010yarbus}, no empirical study has evaluated the optimality of the visual search and free-viewing paradigms in the context of viewer-based object detection. 

To this end, we present \textbf{PET} or Pascal animal classes Eye Tracking database\footnote{The database will be made publicly available upon acceptance of this manuscript.}. PET comprises eye movement recordings compiled for the \textit{bird}, \textit{cat}, \textit{cow}, \textit{dog}, \textit{horse} and \textit{sheep} training+validation (or trainval) sets from the VOC 2012 image set. These six animal-centric classes were chosen from the 20 object classes in VOC2012 owing to the following reasons: (i) Animal classes such as \textit{cats}, \textit{dogs} and \textit{birds} are particularly difficult to detect using traditional supervised learning methods (\eg, deformable parts model) owing to large intrinsic shape and textural variations~\cite{Parkhi11}, and (ii) It would be highly beneficial to incorporate human knowledge to train object detectors for these classes as many psychophysical studies~\cite{Judd_2009} have noted our tendency to instantaneously detect animals (which are both predators and prey).  

A salient aspect of the PET dataset is that it contains eye movements recorded under both \textit{free-viewing} (no task) and \textit{visual search} (task-based) paradigms. In all, eye movements were recorded from a total of 40 users who viewed each image for 2 seconds, so that four gaze patterns are available per image and condition. Comparing high-level eye movement statistics as well as temporal scan paths observed for the two conditions, we observe systematic differences in line with Yarbus' observations~\cite{Yarbus1967}. However, if eye movements are recorded with the \textit{sole objective} of using fixated locations to train object detectors, our analysis suggests that both paradigms are equally suitable for this purpose. Finally, as a utility application, we show how the spatial pooling of fixated eye locations enhances object classification performance for these classes. Through the PET database, we make the following contributions:
\begin{itemize}[noitemsep,nolistsep]
\item[1.] While the fact that task influences eye movement patterns is widely known, the explicit impact of the visual search task on target detection in images (animal-centric VOC classes in our case) has never been empirically studied and quantified. PET represents the first work in this direction, considering approaches that have focused on viewer-centered object detection.
\item[2.] We systematically analyze eye movement behavior in the free-viewing and visual search conditions to show that while the visual search paradigm may in general be more beneficial for target detection, there is little difference between the two paradigms if only the fixated locations are used for subsequent learning.
\item[3.] We show that feature pooling around fixated locations enhances (animal) object classification performance.  
\end{itemize}


The paper is organized as follows. Section~\ref{RW} presents related work, while Section~\ref{EP} describes the experimental protocol employed for compiling the PET data. Section~\ref{SA} compares and contrasts the free-viewing and visual search conditions based on statistical observations, while Section~\ref{ML} discusses how pooling of features extracted around the fixated locations improves classification accuracy for the animal classes. We conclude the paper with key observations in Section~\ref{Con}.

\section{Related work}\label{RW}
Given that humans tend to understand complex scenes by focusing their attentional resources on a small number of \textit{salient} objects~\cite{spainPerona08}, the practice of employing user inputs to interactively understand visual content has been in vogue for sometime now-- the \textit{ESP game}\footnote{\url{en.wikipedia.org/wiki/ESP_game}} is a relevant and popular example. 

While the impact of high-level factors such as cognitive task on eye movements has been extensively studied since the pioneering work of Yarbus~\cite{Yarbus1967}, understanding and predicting the salient content in the scene has only been attempted for over a decade now. While the early works on saliency modeling such as~\cite{Itti1998} hypothesized that our visual attention is guided by low-level image factors such as brightness, contrast and orientation, more recent works~\cite{Judd_2009,ramanathan2010eye} have noted that we are equally likely to focus on semantic entities such as faces, animals and vehicles such as cars. 

The above observations have encouraged direct incorporation of eye movement information for model training in object detection. Two such works that have expressly studied the use of eye movements for improving object detection performance on the Pascal Visual Object Classes (VOC) image set are~\cite{kiwon_cvpr13_gaze} and~\cite{ PapadopoulosCKF14}. In~\cite{kiwon_cvpr13_gaze}, the authors perform several experiments to explore the relationship between image content, eye movements and text descriptions. Eye movements are then used to perform gaze-enabled object detection and produce image annotations.~\cite{PapadopoulosCKF14} uses eye movements for training a model to draw bounding boxes on target objects, upon learning the spatial extent of these objects from fixation and content-based cues.  

However, the point of contention between~\cite{kiwon_cvpr13_gaze} and~\cite{PapadopoulosCKF14} is that~\cite{kiwon_cvpr13_gaze} assumes that natural gaze patterns are already capable of revealing the locations of target objects in the image, while~\cite{PapadopoulosCKF14} explicitly instructs observers to perform a visual search on the images. In contrast, we record eye movements under both paradigms, and analyze if a visual search task explicitly improves user-centric target detection performance. Table~\ref{tab:comp} compares various aspects concerning~\cite{kiwon_cvpr13_gaze,PapadopoulosCKF14} and our work. A detailed explanation of the experimental protocol adopted for PET is as follows.    

%

\begin{table*}[htbp]
\small
\begin{center}
\renewcommand{\arraystretch}{1.8}
\caption{Overview of the three datasets containing eye movement recordings for the VOC image set.} \label{tab:comp}
\begin{tabular}{|c|c|c|c|}
  \hline
	\parbox{0.15\linewidth}{\centering \textbf{Attribute}}&\parbox{0.26\linewidth}{\centering \textbf{SBU GDD}~\cite{kiwon_cvpr13_gaze}}&\parbox{0.26\linewidth}{\centering \textbf{POET}~\cite{PapadopoulosCKF14}}&\parbox{0.26\linewidth}{\centering \textbf{PET} (our)} \\ \hline
\parbox{0.15\linewidth}{\centering \textbf{Objective}}& \parbox{0.26\linewidth}{{Using eye movements and descriptions to improve object detection performance.}}& \parbox{0.26\linewidth}{{Using eye fixations to implicitly annotate bounding boxes for training object detectors.}} & \parbox{0.26\linewidth}{{Using eye movements to improve detection/classification of animal categories in the VOC image set.}}\\ \hline
\parbox{0.15\linewidth}{\centering \textbf{Stimuli}}& \parbox{0.26\linewidth}{{1000 images from 20 object categories (50/class) in VOC2008.}}& \parbox{0.26\linewidth}{{\vspace{0.05in} 6270 images from \textit{cat}, \textit{dog}, \textit{horse}, \textit{cow}, \textit{bicycle}, \textit{motorbike}, \textit{sofa} and \textit{diningtable} classes.}} & \parbox{0.26\linewidth}{{\vspace{0.05in} 4135 images from the \textit{cat}, \textit{dog}, \textit{bird}, \textit{cow}, \textit{horse} and \textit{sheep} classes in VOC2012.}}\\ \hline
\parbox{0.15\linewidth}{\centering \textbf{Task}}& \parbox{0.26\linewidth}{\centering {Free-viewing}}& \parbox{0.26\linewidth}{\centering {Visual search}} & \parbox{0.26\linewidth}{{\textit{Both} free-viewing and visual search}}\\ \hline
\parbox{0.15\linewidth}{\centering \textbf{Number of gaze patterns/image}}& \parbox{0.26\linewidth}{\centering 3} & \parbox{0.26\linewidth}{\centering 5} & \parbox{0.26\linewidth}{{4 each for free-viewing and visual search}}\\ \hline 
\parbox{0.15\linewidth}{\centering \textbf{Stimulus protocol}}& \parbox{0.26\linewidth}{Each image presented for 3 seconds.} & \parbox{0.26\linewidth}{Pairs of images presented until user responds to indicate presence of a target object class.} & \parbox{0.26\linewidth}{{Each image presented for 2 seconds.}} \\ \hline

\end{tabular}
\end{center}
\end{table*}

\section{Materials and Methods}\label{EP}
\textbf{Stimuli and Participants:} 4135 images from the Pascal VOC 2012 dataset~\cite{pascalvoc2012} were selected for the PET study. These images contained one or more instances of the \textit{bird, cat, cow, dog, horse,} and \textit{sheep} categories, and also humans. 2549 images contained \textit{exactly} a single instance of the above classes (also called target classes), while 1586 images contained either multiple instances from the animal classes, or a mix of animals and humans. Considering only images that contained multiple animals, the mean number of animals per image was 3.1$\pm$2.68, which covered a 0.45$\pm$0.28 fraction of the image area (based on bounding box annotations available as part of the VOC dataset). 
A total of 40 university students (18--24 years, 22 males) took part in the experiments. \\

\textbf{Experimental protocol:} Each participant performed the eye-tracking experiment over two sessions spanning about 40 minutes with a short break in-between. They were required to view about 800 images in two blocks, with each image displayed for a duration of 2 seconds and a blank screen displayed in between each image for 500 milliseconds. All participants were instructed to 'free-view' the first block, and asked to 'find all animals in the scene' (visual search) for the second block. The visual search task was always enforced after the free-viewing task to avoid any viewing biases. Also, to minimize boredom, a few \textit{distractor} images which did not contain a single instance of the animal classes were included in the two blocks. The order of the two blocks of images was counterbalanced across a set of subjects, while the images in each block were shown randomly. All images were displayed at $1280\times1024$ resolution on a 17'' LCD monitor placed about 60 cm away from the subjects. Their eye movements were recorded using a Tobii desktop eye tracker, which has a 120 Hz sampling frequency and is accurate to within $0.4^\circ$ visual angle upon successful calibration. \\


\textbf{Pre-processing:} Prior to our analysis, we left out the first fixation on each image, and those fixations with invalid $(x,y)$ coordinates. This resulted in total of 28733 fixations for free viewing, and 29901 fixations for visual search task. Upon pre-processing, for the free-viewing condition, fixation data was available for 3.8$\pm$0.4 users per image, while the number of fixations per image was between 5--45 (mean 24.7$\pm$6.1). For the visual search task, 4$\pm$0.6 gaze patterns were available per image with the number of fixations ranging between 4--52 (mean 22.6$\pm$6.4).


\section{Free-viewing vs visual search}\label{SA}
In this section, we systematically compare gaze behavior in the free-viewing and visual search paradigms and examine if either task benefited viewer-based object detection. 

\begin{figure}[htbp]
\includegraphics[width=8.5cm]{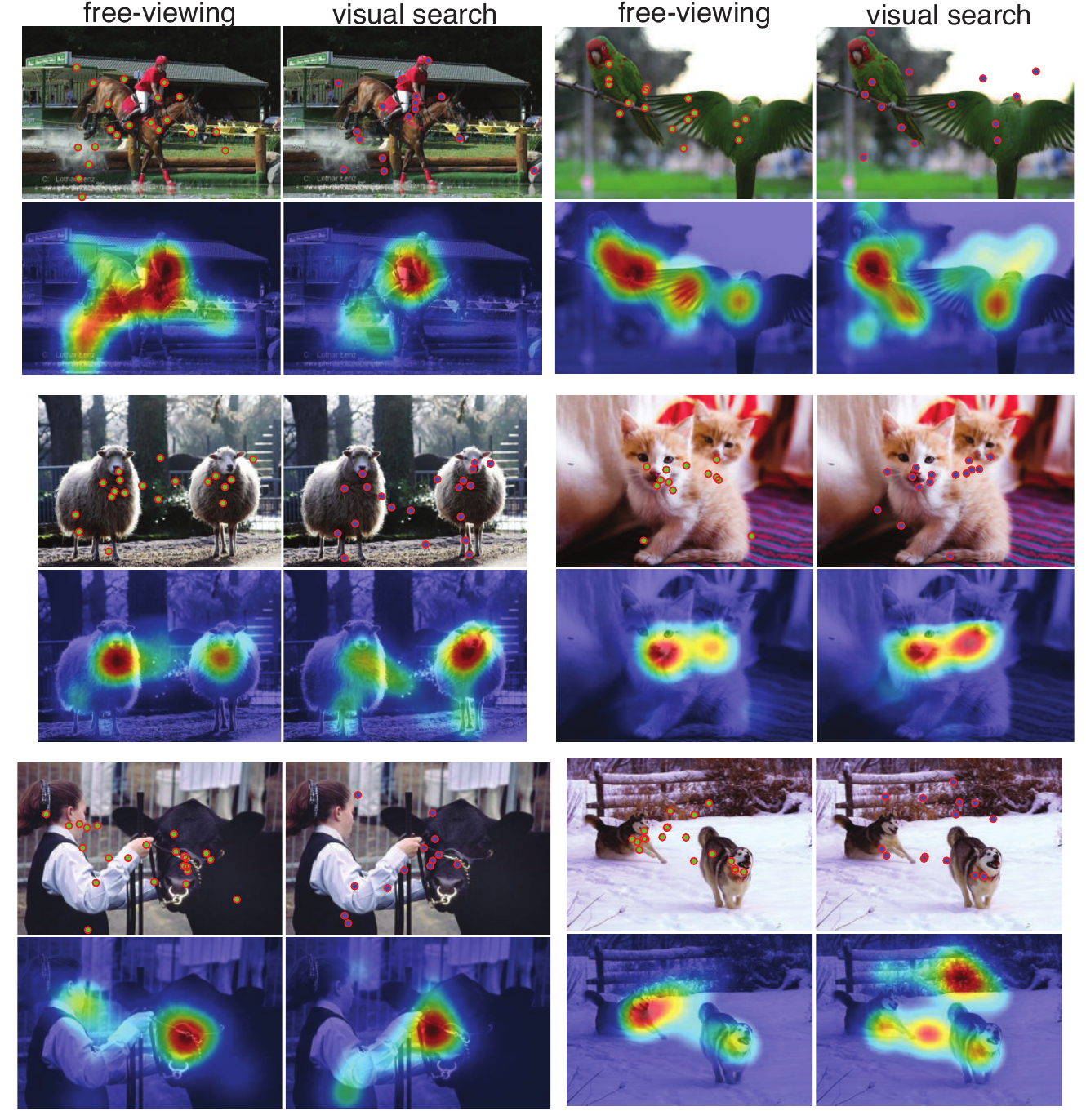}\vspace{-.05in}
\caption{Recorded eye fixations and fixation density maps for the six animal categories considered in PET.}
\label{fig:fixdens}
\end{figure}
\vspace{-.1in}
\subsection{Fixation density maps and overall statistics}

Fixation density maps (or heat maps) qualitatively reveal those regions that are most frequently visited in a scene, and also provide a measure of fixation dispersion in the scene. Fig.~\ref{fig:fixdens} shows fixation density maps for the six animal categories considered in PET. Considering pairs of rows, the top row presents eye fixations made by observers for the free-viewing and visual search tasks, while the bottom row presents the corresponding fixation density maps obtained by convolving fixated scene locations with a Gaussian kernel of bandwidth of $2^\circ$ visual angle. Visual inspection of Fig.~\ref{fig:fixdens} reveals that roughly similar density maps are obtained for both conditions, and that fixations mainly lie on faces.


Bounding box annotations for objects from the 20 VOC classes are available for the trainval sets that are part of Pascal VOC~\cite{pascalvoc2012}. Using bounding box coordinates in images containing \textit{multiple instances} of the target object classes and considering only the \textit{first five fixations} made by each user\footnote{we considered the first five fixations since they are most likely to convey the intent of the viewer.} we determined (i) the proportion of fixations that fell within the bounding boxes for the two conditions-- the proportion was  found to be $0.33\pm0.26$ for both conditions. (ii) the proportion of target objects that had at least one fixation falling on them-- this was again found to be $0.73\pm0.26$  for both conditions. (iii) the time taken by each user to fixate on at least half the number of target objects in the image, termed as \textit{saccadic latency}. Saccadic latency for visual search ($0.40\pm0.34$) was found to be lower than for free-viewing ($0.48\pm0.35$), and a two-sample $t$-test confirmed that this difference was highly significant ($p<0.000001$). (iv) the mean fixation durations per target object in the two conditions-- mean durations were $0.51\pm0.26$ and $0.47\pm0.28$ for the free-viewing and visual search conditions, and the difference was again highly significant ($p<0.000001$).

\begin{figure*}[!ht]
    \centering
    \begin{subfigure}[t]{0.22\textwidth}
        \centering
        \includegraphics[height=1.2in]{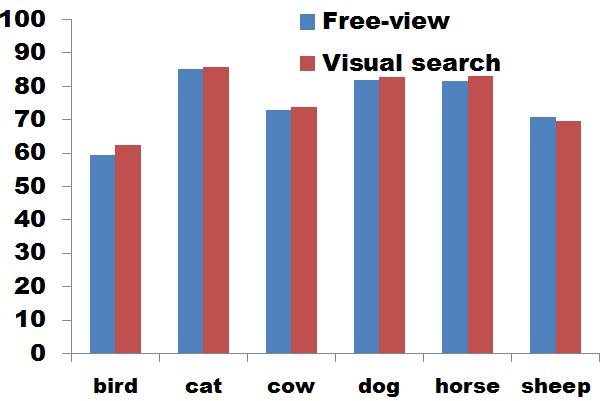}
        \caption{Class-wise distribution of fixation proportions (in \%).}\label{cwfp}
    \end{subfigure}%
    ~
    \begin{subfigure}[t]{0.37\textwidth}
        \centering
        \includegraphics[height=1.2in]{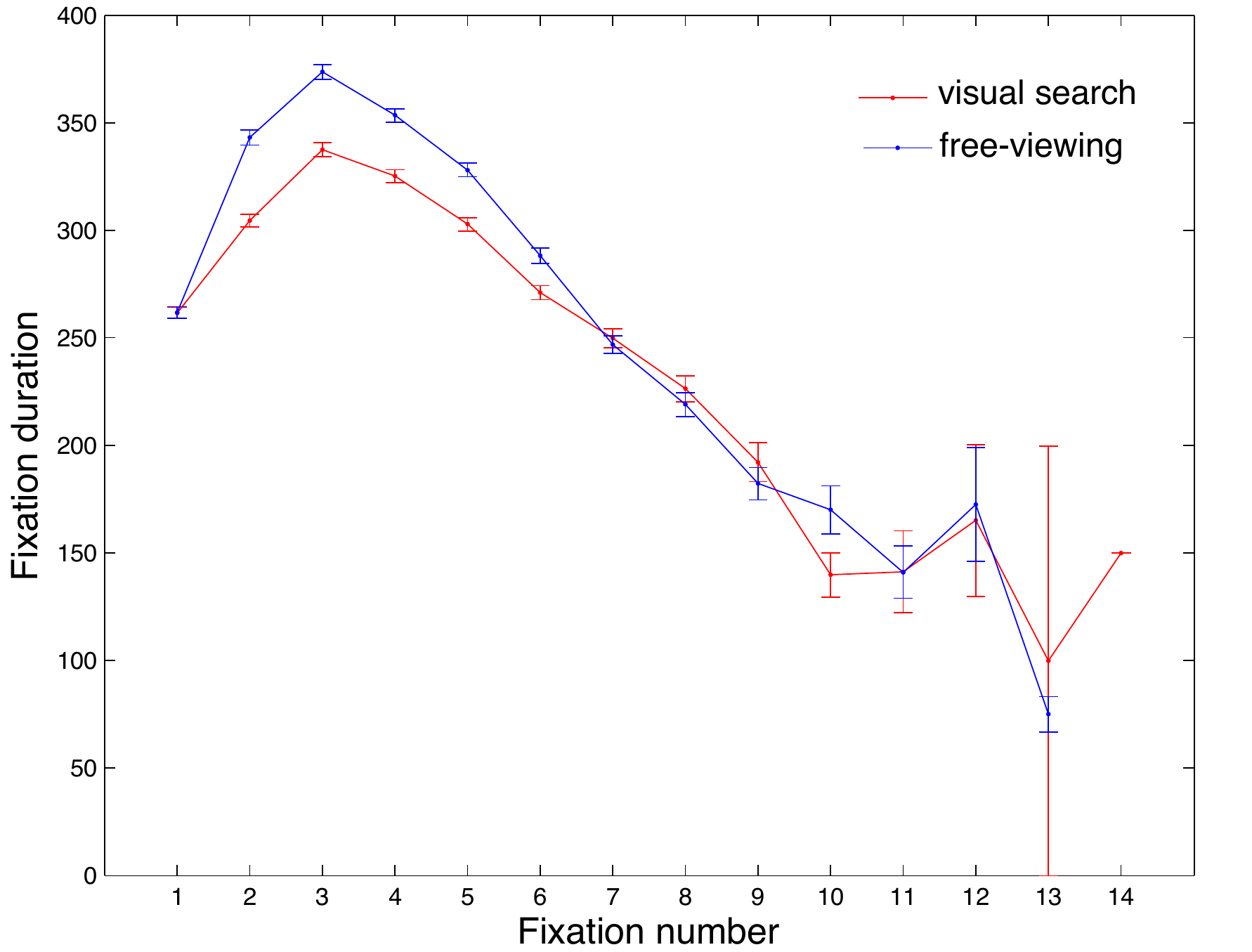}
        \caption{Comparing per-fixation durations ($\mu \pm \sigma$ shown) for visual search and free-viewing conditions}\label{pfdc}
    \end{subfigure}
		~
    \begin{subfigure}[t]{0.37\textwidth}
        \centering
        \includegraphics[height=1.2in]{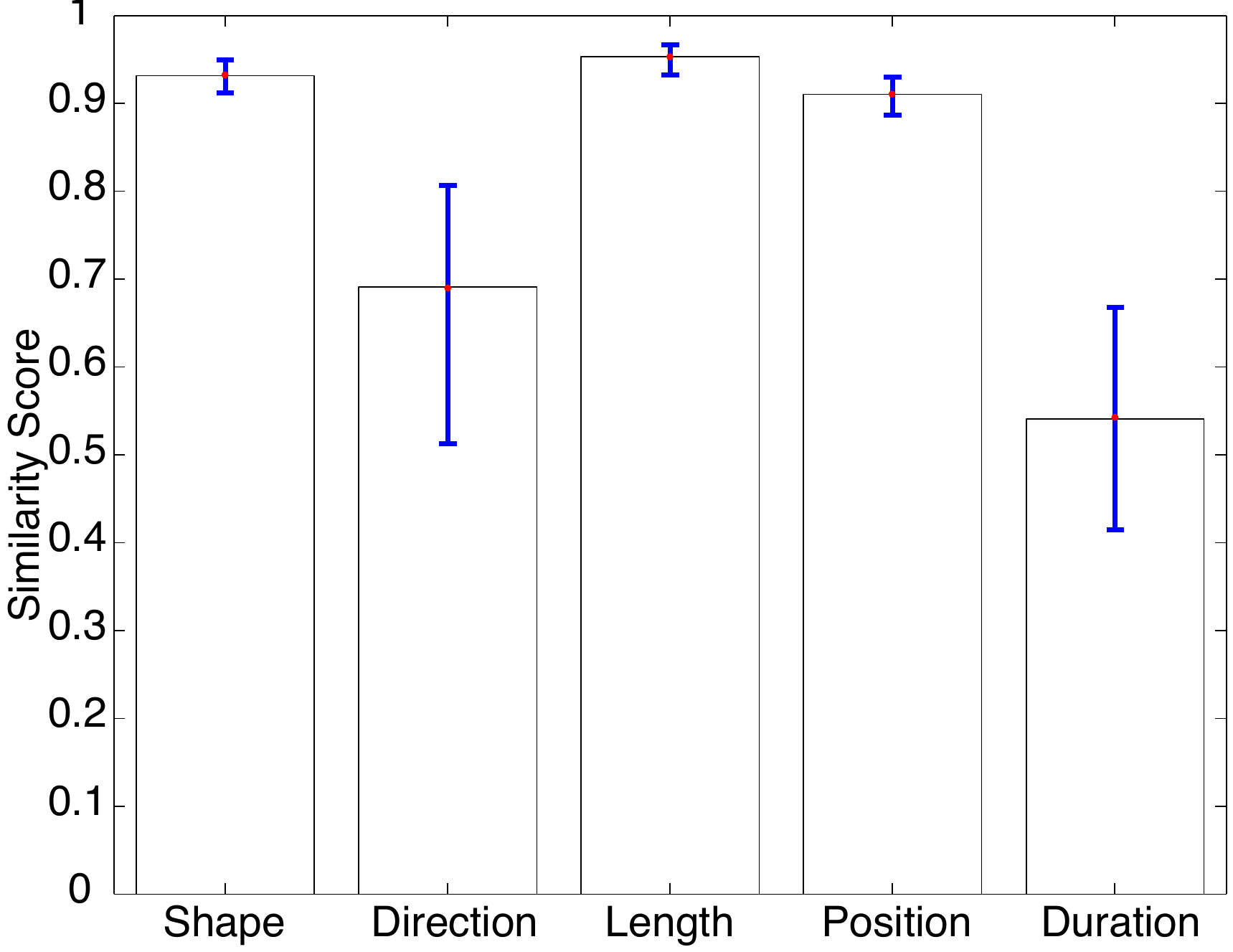}
        \caption{Multimatch similarity scores between free-viewing and visual search tasks.}
				\label{Mmatch}
    \end{subfigure}
		\caption{Free-view vs visual search comparisons (best viewed in color)}
  \end{figure*}

We also computed proportion of fixations falling within the target bounding box for the two conditions for each of the animal classes, to examine if any artifacts affected the overall observed statistics. Fig.~\ref{cwfp} presents the class-wise distribution for both conditions, and shows that the proportions are roughly equal for all of the classes.

\subsection{Per-fixation durations}
Moving on from overall statistics, we now focus on fine-grained analysis of gaze patterns to examine the free-viewing and visual search paradigms. Firstly, we examined the duration of each fixation made by the population of users for the two conditions. As seen from Fig.~\ref{pfdc}, the first few fixations were longer and are followed by progressively shorter fixations. Also, consistent with the overall numbers, per-fixation durations for free-viewing were consistently higher as compared to visual search up to the sixth fixation, while subsequent fixations were similar in duration.



\subsection{MultiMatch and Recurrence Quantification analysis}
In multimatch analysis~\cite{Dewhurst}, the series of fixations made by a viewer is treated as a vector (denoting the scan path). Then, the set of scan paths obtained for two conditions are processed to quantify the inter-conditional differences in terms of saccade shape, length and direction, aligned fixation locations and durations. The algorithm returns a similarity score in the range [0-1]. Figure~\ref{Mmatch} presents the similarity between the above measures characterizing viewing behavior during the free-viewing and visual search tasks (considering all subjects and images). In general, these results show that the gaze behaviors in the free-viewing and visual search conditions are similar in a number of respects. Considerable differences are observed only with respect to saccade direction and fixation duration, for which low similarity scores are obtained. 



\begin{figure}[htbp]
\includegraphics[width=8cm]{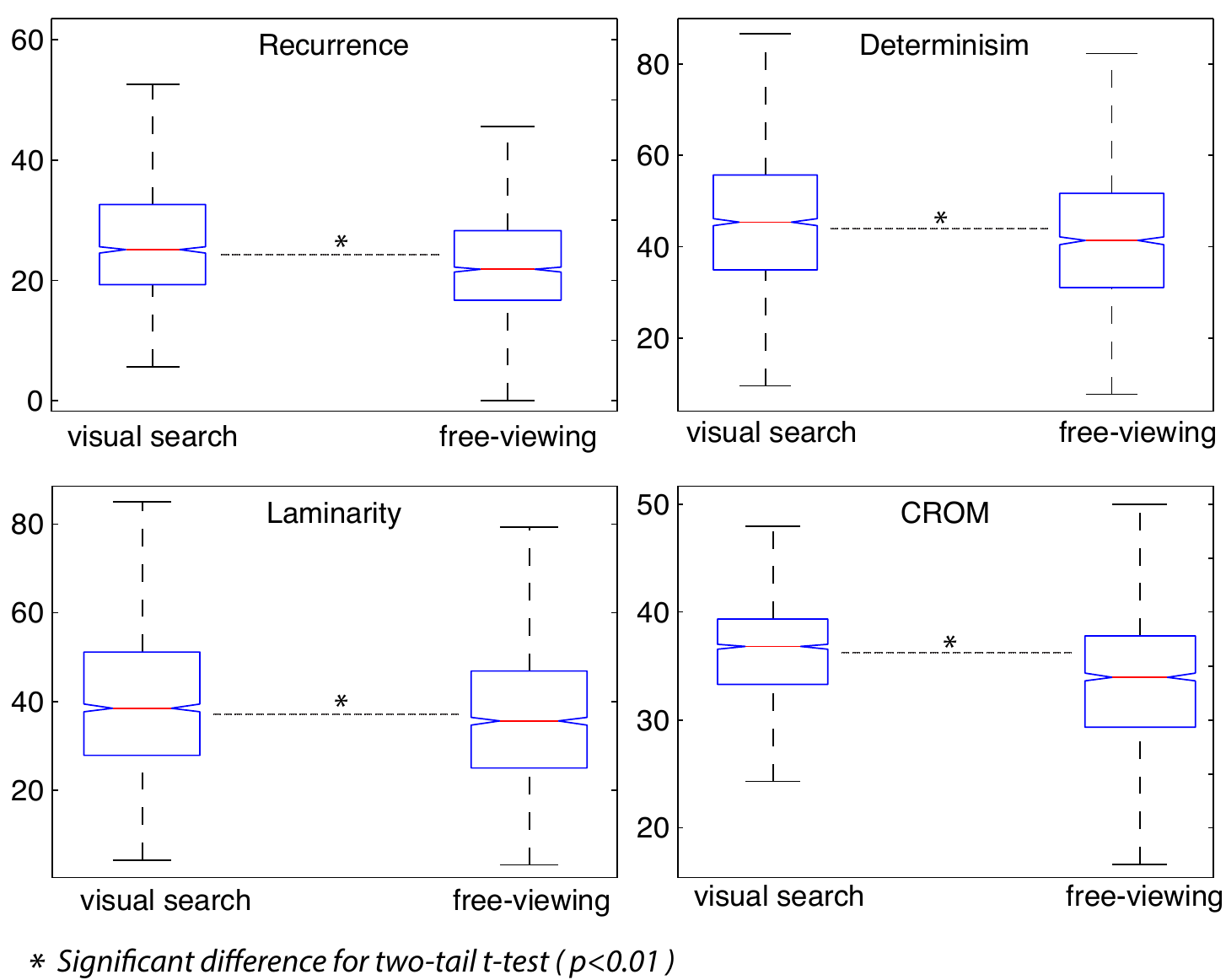}\vspace{-0.02in}
\caption{RQA results for the free-viewing and visual search conditions. Error bars denote unit standard deviation.}
\label{fig:RQA}
\end{figure}

The recurrence quantification analysis (RQA) technique examines the dynamics of a single scan path \cite{Anderson}, and provides a measure of the \textit{compactness} in viewing behavior. The compactness is quantified using measures such as \textit{Recurrence}-- the proportion of fixations that are repeated on previously fixated locations; \textit{Determinism}-- the proportion of recurrent fixations representing repeated gaze trajectories; \textit{Laminarity} denoting the proportion of fixations in a region being repeatedly fixated, and \textit{center of recurrent mass} (CROM), which measures the temporal proximity of recurrent fixations (\ie, time-interval between recurrent fixations). Figure~\ref{fig:RQA} presents the RQA results comparing the visual search and free-viewing conditions. Interestingly, viewing behavior in the visual search scenario was found to be significantly more compact than for free-viewing, in terms of all four measures ($p<0.01$ with two-tailed $t$-tests). 

\subsection{Discussion}
Analysis of the viewing behavior in the free-viewing and visual search tasks reveals that viewers in general, tended to fixate on the target objects quicker (lower saccadic latency) and showed a greater urgency in moving around the scene (lower overall fixation and per-fixation duration) during visual search. The multimatch and recurrence quantification analyses show up differences in terms of saccade direction, and the compactness of fixated locations for the two tasks. The fact that scan paths were found to be more compact during visual search suggests that viewers tended to recurrently traverse (what they perceived as) the informative parts of the scene instead of focusing on peripheral scene details. Based on these observations, we make the following comments regarding the suitability of the free-viewing and visual search tasks for viewer-based target detection. 

\begin{itemize}[noitemsep,nolistsep]
\item[1.] The visual search task appears to motivate viewers to preferentially fixate on designated targets (animals), and traverse the scene with more urgency as compared to free-viewing.
\item[2.] Nevertheless, there is not much to choose between the two paradigms if only the fixated locations are to be considered for model training. The proportion of fixations observed on target objects as well as the proportion of target objects fixated are very similar for both conditions.  
\end{itemize}

\section{Gaze-assisted object classification}\label{ML}

In this section, we show how the eye fixations made by viewers in the target detection task can be useful for enhancing object classification accuracy\footnote{Unlike detection, classification does not localization of the object. We employ implicitly acquired eye fixations to aid the same.}. The bag-of-words model\footnote{\url{http://en.wikipedia.org/wiki/Bag-of-words_model}} is extremely popular in image classification. However, it does not encode any spatial information. Spatial pyramid histogram representation is a more sophisticated approach in this respect, as it includes spatial information for object classification and consists of two steps-- \textit{coding} and \textit{pooling}.

\begin{figure}[htbp]
\includegraphics[width=8.3cm]{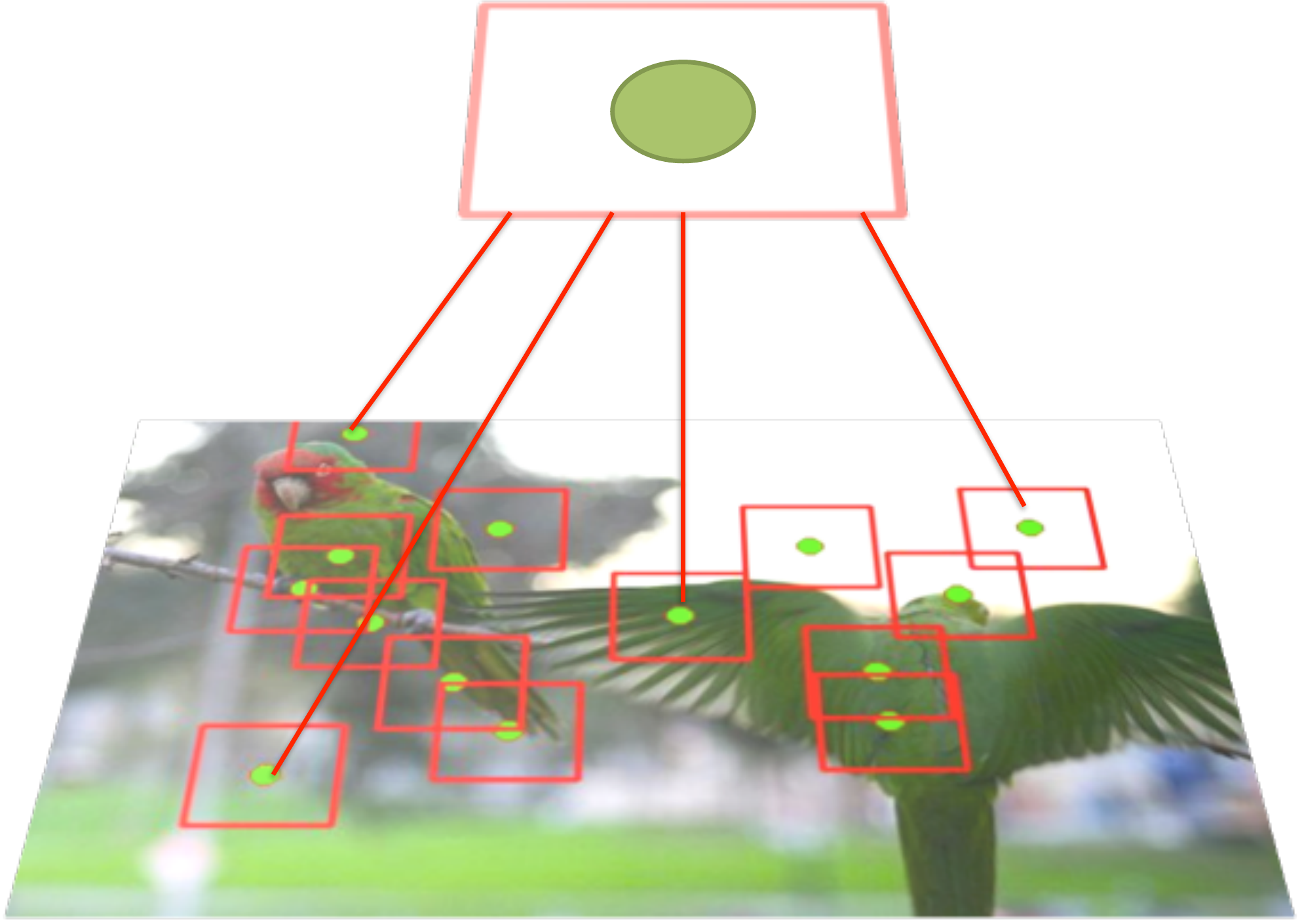}\vspace{-0.02in}
\caption{Feature pooling based on fixated locations: Green dots denote eye fixations. SIFT features are pooled within a window of size 30 $\times$ 30 around the fixated location.}
\label{fig:pool}
\end{figure}

The coding step involves point-wise transformation of descriptors to a representation better adapted to the task. The pooling step combines outputs of several nearby feature detectors to synthesize a local or global bag of features. Pooling is employed to (i) achieve invariance to image transformations, (ii) arrive at more compact representations, and (iii) achieve better robustness to noise and clutter. However, the spatially pooled regions are usually naively defined in literature. Spatial pyramid match~\cite{Yang09CVPR} works by partitioning the image into increasingly finer sub-regions, and computing histograms of local features inside each sub-region. These regions are usually \textit{squares} which are \textit{sub-optimal} for due to the inclusion of unnecessary background information. Given that viewers tend to fixate on meaningful scene regions, the fixated locations can provide a valuable cue regarding the image features to be used for learning. Therefore, in this work, we pool features from regions around the fixated locations instead of sampling from all over the image
Fig.~\ref{fig:pool} illustrates the architecture of our proposed fixation-based feature pooling approach.

Linear spatial pyramid matching with sparse coding~\cite{Yang09CVPR} has successfully used for object classification. Sparse coding has been shown to find succinct representations of stimuli, and model data vectors as a linear combination of a few dictionary codewords. In this paper, sparse coding is adopted for the coding step. We then evaluate the effect of different pooling strategies. Sparse coding is defined as follows:

\begin{equation}
\label{eqn:1}
\begin{array}{l}
\mathop {\min }\limits_{\mathbf{D,C}} {\left\| {\mathbf{X} - \mathbf{C D} } \right\|_F^2 + \lambda _1 {\left\| {\mathbf{C} } \right\|_1 } } \\ \nonumber
s.t. \quad {\mathbf{D_{j \cdot } D_{j }^T} \le 1, \quad \forall j = 1,...,l} \\
\end{array}
\end{equation}

\noindent where $\mathbf{X = [x_1, x_2, ..., x_n]}^T \in R^{d \times n}$, $\mathbf{x_i}$ is the $d$-dimensional feature vector and $n$ is the number of training samples. $\mathbf{D} \in R^{l \times d}$ is an overcomplete dictionary ($l > d $) with $l$ prototypes. $\mathbf{C} \in R^{n \times l}$ corresponds to the sparse representation of $\mathbf{X}$. $\lambda_1$ is the regularization parameter, while $\mathbf{D_{j}}$ denotes the $j$-th row of $\mathbf{D}$. The sparsity constraint prevents the learned dictionary from being arbitrarily large.

Popular pooling strategies are average and max pooling. Average pooling is defined as $\mathbf{p} = \frac{1}{m}\sum\limits_{i = 1}^m {\mathbf{c_i} }$, while max pooling
is defined as $p_k = \max \{ \left| {c_{1k} } \right|,\left| {c_{2k} } \right|,...,\left| {c_{mk} } \right|\}$, where $p_k$ is the $k$-th element of $\mathbf{p}$, $c_{ij}$ is the element at position $(i,j)$ in $\mathbf{C}$. $m$ is the local number of local descriptors in the considered image region.

\textbf{Experimental Results:} Instead of pooling features from a regular spatial pyramid, we pool features around fixated locations. Sparse representations of SIFT features are pooled within a window of size 30 $\times$ 30 pixels around each fixation. Table \ref{tab:VOCclassification} compares the impact of different pooling strategies on animal classification for the PET images. All experiments were repeated five times and average accuracies and standard deviations are reported. A linear SVM classifier is used as in~\cite{Yang09CVPR}. As in \cite{Yang09CVPR}, we observe that max pooling based on sparse codes generally outperforms average pooling on the PET image set. Moreover, eye fixation-based max pooling achieves the best results on 4 of the 6 considered animal classes. More than 3\%  improvement in average classification accuracy is achieved with respect to max pooling using a regular spatial pyramid. Finally, given the compactness of gaze patterns in the visual search task, we achieve slightly better classification accuracy using eye fixations recorded from visual search as compared to free-viewing.
\begin{table*}[t]
\caption{Object classification accuracy with different pooling strategies on PET images.}
\label{tab:VOCclassification}
\centering
\vspace{-0.1in}
\resizebox{0.99\linewidth}{!} {
\begin{tabular}{|c||c|c|c|c|c|c||c|}
\hline
& bird & cat & cow & dog & horse & sheep & avg+std\\ \hline \hline
max pooling \cite{Yang09CVPR} & 0.333 $\pm$ 0.018 & 0.263 $\pm$ 0.058 & 0.220 $\pm$ 0.032 & \textbf{0.522 $\pm$ 0.024} & 0.589 $\pm$ 0.036 & 0.437 $\pm$ 0.045 & 0.394 $\pm$ 0.007 \\ \hline
avg pooling & 0.323 $\pm$ 0.035 & 0.283 $\pm$ 0.026 & 0.222 $\pm$ 0.041 & 0.463 $\pm$ 0.018 & 0.517 $\pm$ 0.044 & 0.402 $\pm$ 0.051 & 0.368 $\pm$ 0.012\\ \hline
max pooling @ eye fixation (visual search) & \textbf{0.357 $\pm$ 0.022} & 0.278 $\pm$ 0.034 & \textbf{0.253 $\pm$ 0.023} & 0.517 $\pm$ 0.016 & \textbf{0.659 $\pm$ 0.021} & \textbf{0.472 $\pm$ 0.031} & \textbf{0.423 $\pm$ 0.011} \\ \hline
avg pooling @ eye fixation (visual search) & 0.346 $\pm$ 0.021 & \textbf{0.291 $\pm$ 0.014} & 0.247 $\pm$ 0.036 & 0.508 $\pm$ 0.009 & 0.547 $\pm$ 0.022 & 0.441 $\pm$ 0.038 & 0.396 $\pm$ 0.021\\ \hline
max pooling @ eye fixation (free-viewing) & 0.348 $\pm$ 0.019 & 0.264 $\pm$ 0.023 & 0.242 $\pm$ 0.019 & 0.499 $\pm$ 0.025 & 0.635 $\pm$ 0.018 & 0.457 $\pm$ 0.028 & 0.408 $\pm$ 0.009 \\ \hline
avg pooling @ eye fixation (free-viewing) & 0.341 $\pm$ 0.017 & 0.251 $\pm$ 0.018 & 0.224 $\pm$ 0.031 & 0.487 $\pm$ 0.012 & 0.526 $\pm$ 0.015 & 0.428 $\pm$ 0.026 & 0.376 $\pm$ 0.014\\ \hline
\end{tabular}
}
\end{table*}

\section{Conclusions}\label{Con}
The presented PET database contains eye movement recordings compiled exclusively for trainval images from the six animal categories in the Pascal VOC 2012 dataset. A salient aspect of PET is that it contains eye movements recorded under both \textit{free-viewing} and \textit{visual search} conditions. Systematic comparison of gaze patterns for the two conditions suggests that while visual search appears to motivate the viewer better to perform target detection, target objects are fixated in equal measure under both conditions. Object classification accuracy is found to improve by pooling SIFT features around fixated locations, and pooling features around fixations acquired during visual search is more beneficial given the compactness of gazed locations observed for this condition as compared to free-viewing.
\bibliographystyle{IEEEbib}
\bibliography{icme2015template}

\end{document}